\pdfoutput=1 
\documentclass{article}

\usepackage{subcaption}
\usepackage{wrapfig}
\usepackage{tikz} 
\usetikzlibrary{calc} 
\usetikzlibrary{arrows.meta} 
\usetikzlibrary{trees}

\usepackage{longtable}
\usepackage{colortbl}
\usepackage{booktabs}
\usepackage{adjustbox}
\usepackage{array} 
\usepackage{tabularx}
\usepackage{longtable,array,multirow}
\usepackage{multirow}
\usepackage{etoolbox}
\usepackage{xparse}

\usepackage[boxed, linesnumbered]{algorithm2e}
\usepackage{hyperref}
\hypersetup{colorlinks,allcolors=black}

\usepackage{enumitem}
\setlist[itemize]{noitemsep, topsep=0pt}

\usepackage[acronym, toc=false, accsupp]{glossaries-extra}

\glsdisablehyper

\setabbreviationstyle[acronym]{long-short}
\setabbreviationstyle{long-short}

\newacronym{ate}{ATE}{average treatment effect}
\newacronym{mle}{MLE}{maximum likelihood estimation}
\newacronym{mab}{MAB}{multi-armed bandit}
\newacronym{cmab}{CMAB}{contextual multi-armed bandit}
\newacronym{rar}{RAR}{response-adaptive randomization}
\newacronym{ci}{CI}{confidence interval}
\newacronym{cara}{CARA}{covariate-adjusted response-adaptive}
\newacronym{ncworks}{NCW4H}{North Carolina Works for Health}
\newacronym{best}{BEST}{Biomarkers for Evaluating Spine Treatments}
\newacronym{act}{ACT}{acceptance and commitment therapy}
\newacronym{ebem}{EBEM}{evidence-based exercise and manual therapy}
\newacronym{esc}{ESC}{enhanced self-care}
\newacronym{ae}{AE}{adverse event}
\newacronym{bacpac}{BACPAC}{Back Pain Consortium}
\newacronym{clbp}{cLBP}{chronic low back pain}
\newacronym{dtr}{DTR}{dynamic treatment regime}
\newacronym{odtr}{oDTR}{optimal dynamic treatment regime}
\newacronym{niams}{NIAMS}{National Institute of Arthritis and Musculoskeletal and Skin Diseases}
\newacronym{nih}{NIH}{National Institutes of Health}
\newacronym{peg}{PEG}{{Pain, Enjoyment of Life, and General Activity}}
\newacronym{pgic}{PGIC}{Patient Global Impression of Change}
\newacronym{sae}{SAE}{serious adverse event}
\newacronym{sap}{SAP}{statistical analysis plan}
\newacronym{smart}{SMART}{sequential, multiple-assignment, randomized trial}
\newacronym{heal}{HEAL}{Helping to End Addiction Long-term}
\newacronym{car}{CAR}{covariate-adaptive randomization}
\newacronym{mri}{MRI}{magnetic resonance imaging}
\newacronym{mrs}{MRS}{magnetic resonance spectroscopy}
\newacronym{pro}{PRO}{patient reported outcome}
\newacronym{snri}{SNRI}{serotonin–norepinephrine reuptake inhibitor}
\newacronym{immpact}{IMMPACT}{{Initiative on Methods, Measurement, and Pain Assessment in Clinical Trials}}
\newacronym{nimhd}{NIMHD}{The National Institute of Minority Health and Health Disparities}
\newacronym{sdoh}{SDOH}{social determinants of health}
\newacronym{sed}{SED}{socioeconomically disadvantaged}
\newacronym{rct}{RCT}{randomized controlled trial}
\newacronym{il}{IL}{individual level}
\newacronym{cl}{CL}{cluster level}
\newacronym{el}{EL}{employer level}
\newacronym{sw}{SW}{stepped wedge}
\newacronym{swd}{SWD}{stepped wedge design}
\newacronym{swt}{SWT}{stepped wedge trial}
\newacronym{gee}{GEE}{generalized estimating equations}
\newacronym{consort}{CONSORT}{consolidated standards of reporting trials}
\newacronym{dp}{DP}{design pattern}
\newacronym{po}{PO}{potential outcome}
\newacronym{cm}{CM}{completeness matrix}
\newacronym{crt}{CRT}{cluster-randomized trial}
\newacronym{crd}{CRD}{cluster-randomized design}
\newacronym{cdpp}{CDPP}{Chronic Disease Prevention Program}
\newacronym{mli}{MLI}{multi-level intervention}
\newacronym{mlest}{MLE}{maximum likelihood estimator}
\newacronym{ols}{OLS}{ordinary least squares}
\newacronym{er}{ER}{equal randomization}
\newacronym{ipw}{IPW}{inverse probability weighting}
\newacronym{aipw}{AIPW}{augmented inverse probability weighting}
\newacronym{edtr}{eDTR}{embedded dynamic treatment regime}
\newacronym{cs}{CS}{computer science}
\newacronym{igw}{IGW}{inverse-gap weighting}
\newabbreviation{dss}{DSS}{Division of Social Services}
\newabbreviation{mitt}{mITT}{modified intention-to-treat}
\newabbreviation{xai}{XAI}{explainable aritificial intelligence}
\newabbreviation{vim}{VIM}{variable importance metric}
\newabbreviation{mar}{MAR}{missing at random}
\newabbreviation{mi}{MI}{multiple imputation}
\newabbreviation{iml}{iml}{interpretable machine learning}

\title{Statistical Design and Rationale of the Biomarkers for Evaluating Spine Treatments (BEST) Trial}

\hypersetup{
pdftitle={Statistical Design and Rationale of the Biomarkers for Evaluating Spine Treatments (BEST) Trial},
pdfsubject={stat.AP},
pdfauthor={John~Sperger},
pdfkeywords={cLBP, dynamic treatment regime, precision medicine, sequential multiple-assignment randomized trial},
}

\newbool{isarxiv}

\newbool{isct}

\booltrue{isarxiv}

\ifbool{isarxiv}{\boolfalse{isct}
}{\booltrue{isct}}

\ifbool{isarxiv}{
    \usepackage{arxiv}
    \usepackage{natbib}
    \usepackage{glossaries-accsupp}

    \renewcommand{\shorttitle}{Statistical Design of the BEST Trial}
    \usepackage{authblk}
    \setlength{\affilsep}{0em}
    \newbox{\orcid}\sbox{\orcid}{\includegraphics[scale=0.06]{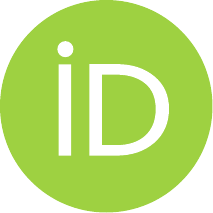}}

    \author[1, 18]{%
    	\href{https://orcid.org/0000-0002-2273-5353}{\usebox{\orcid}\hspace{1mm}John Sperger, PHD\thanks{Corresponding author; direct correspondence to \texttt{jsperger@live.unc.edu}}}%
    }
    \author[2]{Kelley M. Kidwell, PhD}
    \author[3]{Matthew C. Mauck, MD PhD}
    \author[4]{Beibo Zhao, PhD}
    \author[4]{Kevin J. Anstrom, PhD}
    \author[4]{Anna Batorsky, DrPH}
    \author[5]{Timothy S. Carey}
    \author[6]{Daniel J. Clauw, MD}
    \author[7,19]{Nikki L. B. Freeman, PhD}
    \author[8]{Carol M. Greco, PhD}
    \author[1]{Anastasia Ivanova, PhD}
    \author[9]{Sara Jones Berkeley, MPH PhD}
    \author[10]{Samuel A. McLean, MD}
    \author[11]{Matthew A. Psioda, PhD}
    \author[4]{Bryce Rowland, PhD}
    \author[12,15,16,17]{Gwendolyn A. Sowa, MD PhD}
    \author[13]{Ajay D. Wasan, MD MSc}
    \author[14]{Joshua P Zitovsky, PhD}
    \author[1]{Michael R. Kosorok, PhD}
    
    \affil[1]{Department of Biostatistics, Gillings School of Global Public Health, University of North Carolina at Chapel Hill}
    \affil[2]{Department of Biostatistics, School of Public Health, University of Michigan} 
    \affil[3]{Department of Anesthesiology, University of North Carolina at Chapel Hill}
    \affil[4]{ Collaborative Studies Coordinating Center, Department of Biostatistics, Gillings School of Global Public Health, University of North Carolina at Chapel Hill} 
    \affil[5]{Department of Medicine, University of North Carolina at Chapel Hill}
    \affil[6]{Department of Anesthesiology, Chronic Pain \& Fatigue Research Center, University of Michigan, Ann Arbor}
    \affil[7]{Department of Biostatistics and Bioinformatics, Duke University}
    \affil[8]{Department of Psychiatry, University of Pittsburgh School of Medicine}
    \affil[9]{Department of Epidemiology, Gillings School of Global Public Health, University of North Carolina at Chapel Hill}
    \affil[10]{Department of Psychiatry, University of North Carolina at Chapel Hill}
    \affil[11]{Statistics and Data Science Innovation Hub, Biostatistics, GSK}
    \affil[12]{Department of Physical Medicine and Rehabilitation, University of Pittsburgh School of Medicine}
    \affil[13]{Departments of Anesthesiology and Perioperative Medicine; and Psychiatry, University of Pittsburgh School of Medicine}
    \affil[14]{Amazon Postdoctoral Scientist}
    \affil[15]{Ferguson Laboratory for Orthopaedic and Spine Research, Bethel Musculoskeletal Research Center, Department of Orthopaedic Surgery, University of Pittsburgh School of Medicine}
    \affil[16]{Department of Bioengineering, University of Pittsburgh Swanson School of Engineering}
    \affil[17]{Clinical and Translational Science Institute, University of Pittsburgh}
    \affil[18]{Carolina Pragmatic Trials Laboratory, North Carolina Translational and Clinical Sciences Institute, UNC School of Medicine}
    \affil[19]{Duke Clinical Research Institute}
       
    \newenvironment{funding}[1]%
    {\subsection*{\normalsize\bfseries Funding}\begin{refsize}\noindent #1}%
    {\end{refsize}}

    {\subsection*{\normalsize\bfseries Supplemental material}\begin{refsize}\noindent #1}%
    {\end{refsize}}
     
    \newenvironment{dci}[1]%
    {\subsection*{\normalsize\bfseries Declaration of conflicting interests}\begin{refsize}\noindent #1}%
    {\end{refsize}} 
    
    \newenvironment{trial}[1]%
    {\subsection*{\normalsize\bfseries Trial registration}\begin{refsize}\noindent #1}%
    		{\end{refsize}}
    \newenvironment{ethics}[1]%
    {\subsection*{\normalsize\bfseries Trial registration}\begin{refsize}\noindent #1}%
    		{\end{refsize}}
}

\ifbool{isct}{
    \runninghead{Sperger \textit{et al.}}

    \author{John Sperger, PhD\affilnum{1},
    Kelley M. Kidwell, PhD\affilnum{2}, 
    Matthew C. Mauck, MD PhD\affilnum{3},
    Beibo Zhao, PhD\affilnum{4},
    Kevin J. Anstrom, PhD\affilnum{4},
    Anna Batorsky, DrPH\affilnum{4},
    Timothy S. Carey\affilnum{5},
    Daniel J. Clauw, MD\affilnum{6},
    Nikki L.B. Freeman, PhD\affilnum{13}
    Carol M. Greco, PhD\affilnum{7},
    Anastasia Ivanova, PhD\affilnum{1},
    Sara Jones Berkeley, MPH PhD\affilnum{8},
    Matthew A. Psioda, PhD\affilnum{9},
    Bryce Rowland, PhD\affilnum{4},
    Gwendolyn A. Sowa, MD PhD\affilnum{10},
    Ajay D. Wasan, MD MSc\affilnum{11},
    Joshua P Zitovsky, PhD\affilnum{12},
    Michael R. Kosorok, PhD\affilnum{1}}
    
    \affiliation{\affilnum{1}Department of Biostatistics, Gillings School of Global Public Health, University of North Carolina at Chapel Hill \\
    \affilnum{2}Department of Biostatistics, School of Public Health, University of Michigan \\ 
    \affilnum{3}Department of Anesthesiology, University of North Carolina at Chapel Hill \\
    \affilnum{4} Collaborative Studies Coordinating Center, Department of Biostatistics, Gillings School of Global Public Health, University of North Carolina at Chapel Hill \\ 
    \affilnum{5}Department of Medicine, University of North Carolina at Chapel Hill\\
    \affilnum{6}Department of Anesthesiology, Chronic Pain \& Fatigue Research Center, University of Michigan, Ann Arbor\\
    \affilnum{7}Department of Psychiatry, University of Pittsburgh School of Medicine\\
    \affilnum{8}Department of Epidemiology, Gillings School of Global Public Health, University of North Carolina at Chapel Hill \\
    \affilnum{9}Statistics and Data Science Innovation Hub, Biostatistics, GSK\\
    \affilnum{10}Department of Physical Medicine and Rehabilitation, University of Pittsburgh School of Medicine; Ferguson Laboratory for Orthopaedic and Spine Research, Bethel Musculoskeletal Research Center, Department of Orthopaedic Surgery, University of Pittsburgh School of Medicine; Department of Bioengineering, University of Pittsburgh Swanson School of Engineering; Clinical and Translational Science Institute, University of Pittsburgh \\
    \affilnum{11}Departments of Anesthesiology and Perioperative Medicine; and Psychiatry, University of Pittsburgh School of Medicine \\
    \affilnum{12}Amazon Postdoctoral Scientist \\
    \affilnum{13}Department of Biostatistics and Bioinformatics, Duke University
    }
    
    \corrauth{John Sperger, 
    Department of Biostatistics, The University of North Carolina at Chapel Hill, 
    Chapel Hill, NC 27599-7420, USA.}
    
    \email{jsperger@live.unc.edu}
    
    \keywords{cLBP, dynamic treatment regime, precision medicine, sequential multiple-assignment randomized trial}

    {\subsection*{\normalsize\sagesf\bfseries Trial registration}\begin{refsize}\noindent #1}%
    		{\end{refsize}}
    \newenvironment{ethics}[1]%
    {\subsection*{\normalsize\sagesf\bfseries Trial registration}\begin{refsize}\noindent #1}%
    		{\end{refsize}}
}

\ifbool{isarxiv}{
	
	\newcommand{\symbf}{\boldsymbol}

	\newcommand{\symup}{\mathrm}
}

\DeclareMathOperator*{\argmin}{argmin}
\DeclareMathOperator{\E}{{\rm I\kern-.3em E}}
\DeclareMathOperator{\prob}{{\mathrm{Pr}}}
\newcommand{\indfun}{{\symbf{1}}}

\newcommand{\perfcutoff}{\delta}
\newcommand{\err}{\epsilon}
\newcommand{\covarrv}{X}
\newcommand{\armrv}{X}
\newcommand{\resprv}{Y}
\newcommand{\histrv}{H}
\newcommand{\timeindex}{j} 
\newcommand{\pout}{\resprv^{*}}
\newcommand{\obsindex}{t}
\newcommand{\armindex}{k}
\newcommand{\armindexprime}{\armindex^{\prime}}
\newcommand{\maxarmindex}{\MakeUppercase{\armindex}}
\newcommand{\maxobsindex}{N}

\newcommand{\val}{{\symup{V}}}

\newcommand{\pol}{\pi}
\newcommand{\optpol}{\pol^{*}}

\newcommand{\polhat}{\widehat{\pol}}
\newcommand{\polhatn}{\polhat_{n}}
\newcommand{\polhatnx}{\polhatn(X)}

\newcommand{\minimprob}{\rho}
\newcommand{\minimweight}{\omega}
\newcommand{\minimarmweight}{\alpha}
\newcommand{\minimfunc}{g}
\newcommand{\minimdim}{p}

\begin{document}
\maketitle 

\setlength{\parskip}{5pt plus2pt minus2pt}

\vfill
\section*{Abstract} 

Background/Aims: Chronic low back pain (cLBP) is a widely prevalent condition with potentially severe impacts on day-to-day functioning and quality of life. A variety of evidence-based treatment modalities exist, but all approaches have modest average treatment effects—potentially due to individual variation in treatment response and the diverse etiologies of cLBP. The Biomarkers for Evaluating Spine Treatments (BEST) trial was designed to investigate the potential for using patient features including biomarkers and phenotypic measurements to inform cLBP treatment.

Methods: We provide an overview of the BEST trial and detail the statistical rationale for key study design decisions. We discuss the salient features of the scientific and clinical context of cLBP treatment that motivated the study design and planned analysis, describe modifications to existing methods that were needed because of the setting, and present simulation studies that were used to inform the study design.
 
Results: The BEST trial was designed as a multi-site sequential, multiple-assignment randomized trial (SMART) of four cLBP treatment modalities. The primary objective was to inform a precision medicine approach to the treatment of cLBP by estimating the optimal treatment or combination of treatments based on patient features and individual patient's response to the initial treatment. The primary analysis includes estimating an interpretable dynamic treatment regime for recommending treatments based on patient features and investigating the subgroups that respond best to each one of the treatment modalities. The trial aimed to enroll 630 protocol completers.

Broad eligibility criteria were used to improve the generalizability of study results. Notably, participants could be excluded from one of the four interventions and still be eligible for the study. We modified the minimization method for reducing covariate imbalance to allow for these exclusions.

Conclusion: The BEST trial may pave the way for a more evidence-based approach to individualizing cLBP treatment in clinical practice. The trial collected an extensive set of biomarker and phenotypic measures that may lead to identification of potential treatment mechanisms. To our knowledge it is one of the largest SMARTs focused on decision making to date, and the largest in cLBP.
\ifbool{isarxiv}{
  \keywords{cLBP \and dynamic treatment regime \and precision medicine \and sequential multiple-assignment randomized trial}
}

\clearpage

\section{Introduction}\label{best:introduction}

\Gls{clbp}, defined as pain lasting at least three months with pain occurring most days, affects 10-20\% of US adults and is associated with substantial social and economic burden\citep{hoy2014Global, deyo2015Report}. Treatment planning for 
\gls{clbp} patients has been hampered by a dearth of evidence to support informed decision-making. While multiple treatment modalities have demonstrated efficacy, the average improvement is generally modest, with substantial heterogeneity in patient responses \citep{von1996Course, hirase2021Duloxetine, chou2009Interventional, chou2017Systemic, qaseem2017Noninvasive}. Additionally, there is a lack of directly comparative evidence and studies investigating how these treatments interact with the diverse etiologies of \gls{clbp}. This absence of individualized patient data inhibits identifying patient characteristics that could aid clinicians in developing personalized treatment plans for \gls{clbp} based on each patient's unique features \citep{nihhealinitiative2018RFAAR19027, mauck2023Back}. 

The \gls{best} trial was designed to address this gap. The \gls{best} trial (\texttt{clinicaltrials.gov} ID NCT05396014, Protocol No. 00057948)\label{spirit:registration} is a multi-site, open-label, two-stage \gls{smart} developed as an essential component of the \gls{bacpac} research program. A highly collaborative, patient-centered translational program, \gls{bacpac} is sponsored by
\gls{niams} as part of the \gls{heal} initiative launched by \gls{nih}\label{spirit:funding}. It focuses on developing and testing innovative approaches and technological developments for comprehensive clinical assessment and personalized treatment planning through a combination of cohort studies and clinical trials. The overall structure, research goals, and implementations of \gls{bacpac} have been reported previously \citep{mauck2023Back, carter2023back}. 

The primary objective of the \gls{best} trial was to inform a precision medicine approach to the treatment of \gls{clbp}, central to \gls{bacpac}'s mission, by estimating the optimal treatment or combination of treatments based on patient features and individual patient's response to the initial treatment. \Gls{best} utilized a two-stage \gls{smart} design to provide the highest quality data to address this objective. The trial operationalized this objective through the concept of \gls{dtr}, a framework of decision rules guiding treatment decisions over time based on a patient’s history, and subsequently, the estimation of the \gls{odtr}, an algorithm that specifies treatment rules to maximize treatment effectiveness. Patient subgroups with the most
favorable responses to each treatment could also be identified.

Estimating the \gls{odtr} addressed the following key clinical questions for \gls{clbp} treatment: the maximum potential benefit of incorporating patient characteristics into treatment decisions; the patient features most important for predicting treatment response; the best way to recommend treatments based on patient features to maximize expected improvement; and the development of both comprehensive and clinically interpretable treatment guidelines that account for individual characteristics in selecting initial treatment and guiding subsequent treatment decisions based on initial response. To our knowledge, the \gls{best} trial was the first large-scale \gls{smart} trial with \gls{dtr} estimation as its  objective. Additionally, the \gls{best} trial provided a uniquely comprehensive dataset through an extensive range of biological and phenotypic measurements collected from all participants, offering invaluable insights for the discovery of patient features that predict treatment response and incorporating them into effective, individualized treatment recommendations. 

This manuscript discusses the statistical considerations that informed the design of the \gls{best} trial. \ifbool{isarxiv}{\cite{mauck2025Design}(under review)}{Mauck et al. (under review)\cite{mauck2025Design} } details the study population, selected interventions, data collection and management, assessments, and assessment strategy. Following a brief overview of the study and its objective, we highlight key features of its \gls{smart} design, detail the planned  analysis, present the simulations used to determine sample size and evaluate the minimization-based randomization strategy for reducing imbalances, and discuss the trial's potential impact and limitations.

\section{Study Overview and Objective} \label{best:design-overview}

\subsection{Overview}\label{best:overview}

In the \gls{best} trial, four evidence-based treatments/interventions for \gls{clbp} were evaluated: (1) \gls{esc}, (2) \gls{act}, (3) duloxetine and (4) \gls{ebem}. A panel of \gls{clbp} experts, informed by the biopsychosocial model of pain \citep{engel1977Need, gatchel2007Biopsychosocial}, selected these treatments/interventions based on the strength of the existing evidence in the literature, clinical judgment, and the diversity of the treatment modalities. In this manuscript, we use the terms ``treatments'' and ``interventions'' interchangeably, as the \gls{best} trial evaluates both medical treatments and behavioral interventions. This included harmonized \gls{bacpac} consortium-wide data elements\citep{batorsky2023back}. 

Briefly, each participant in the \gls{best} trial completed an initial screening call and enrollment visit, followed by a 2-week run-in period, two consecutive 12-week treatment periods, and a 12-week post-treatment follow-up. In the first treatment period (stage one), participants were randomly assigned to one of the study interventions. For the second treatment period (stage two), based on their response to the initial treatment, participants either (1) continued with the initial intervention, (2) augmented the initial intervention with an additional randomly assigned intervention, or (3) switched to a new randomly assigned intervention. The exception was for participants who began with \gls{esc}; those not continuing ESC could only augment their treatment since this intervention, by definition, resulted in strategies that remained available to the participant and could not be fully discontinued. Assessments and activities were conducted prior to and at the start of stage one (Baseline/Visit 0), as well as at key timepoints afterward: at the 6-week (midpoint of stage one), the 12-week (end of stage one, Visit 1), the 18-week (midpoint of stage two), the 24-week (end of stage two, Visit 2), and the 36-week (12 weeks post stage two) timepoint.
 
 The  endpoint of the trial was the 24-week \gls{peg}, a validated composite measure assessing pain intensity and its interference with enjoyment of life and general activity over the past week. This measure is based on three patient-reported outcomes on a 0–10 scale: pain intensity, interference with daily activities, and impact on enjoyment of life \citep{krebs2009Development}. Secondary endpoints included additional 24-week measurements of pain interference, opioid use, physical function, depression, anxiety, sleep quality, and sleep duration. 

\subsection{Objective}\label{best:aims}

The primary objective of the \gls{best} trial was to inform a precision medicine approach to the treatment of \gls{clbp} by estimating the optimal treatment or combination of treatments based on patient features and the individual patient's response to the initial treatment. This objective was motivated by the observation that, while multiple, highly evidence-based treatments exist for \gls{clbp}, the average improvement in response to these treatments is generally modest with some patients experiencing marked improvement while others experience no improvement \citep{mauck2022evidence,george2020transforming}. In statistical terms, these treatments show moderate average treatment effects across the population, but there is notable heterogeneity based on patient characteristics. Thus, there is significant interest in identifying patient features, including biomarkers and phenotypic measures, associated with optimal response to specific treatments. This precision medicine approach aims to match the right treatment to the right patient, ultimately leading to greater improvements in \gls{clbp} outcomes.

Given that some patients may not improve with initial treatment or may experience recurring pain over time, a \gls{clbp} patient’s treatment plan often evolves, with treatment order potentially affecting effectiveness. Optimizing outcomes may involve tailoring the sequence of treatments to patient characteristics and their response to prior treatments. For example, patients with low self-efficacy might benefit most from addressing this through \gls{act} before trying other treatments. Those who respond favorably to \gls{act} may continue with it, while those with a partial response may benefit from \gls{esc}, and non-responders to \gls{act} might achieve better outcomes with duloxetine. This consideration shaped the study objective, leading to a focus on identifying optimal treatment combinations that adapt subsequent treatments based on patient responses to the initial treatment, rather than simply comparing individual treatments.

We operationalized the study objective of informing a precision medicine approach to \gls{clbp} into the statistical task of estimating \glspl{dtr} for maximizing the improvement in the primary endpoint of 24-week \gls{peg} for the primary analysis. A \gls{dtr} is a collection of functions that map patients to recommended treatments based on the individual's characteristics; when decisions are made over time these characteristics may include their response to earlier treatments and evolving characteristics. These \glspl{dtr} will be estimated using model ensemble and interpretable machine learning methods. The treatment or combination of treatments that maximizes effectiveness for treating \gls{clbp}, the \gls{odtr}, can be determined from the candidate \glspl{dtr}. The \gls{odtr} identifies the best treatment for a patient based on their features, and this will be complemented by investigating the subgroups of patients that respond best to each treatment modality as part of the primary analysis. 

\section{SMART Design}\label{design}

Facilitating a precision medicine approach to \gls{clbp} requires a trial design capable of informing clinical decision making. Decision support is a distinct statistical goal from statistical inference that requires high-quality data to avoid relying on strong assumptions \citep{wang2012Evaluation, tsiatis2019Dynamic, sperger2020Future}. The study's objective makes the \gls{smart} design particularly suitable, as it accommodates multiple treatments and treatment adaptations based on patient response \citep{wang2012Evaluation, tsiatis2019Dynamic}. Washout periods are not typically part of routine care, and treatment sequencing or ordering effects may be clinically important but can not be estimated when using a crossover
design. Additionally, allowing patients to intensify treatment in stage two by augmenting their first-line treatment with an additional intervention can be a cost-efficient alternative to full factorial designs. Twelve sites participated in the trial to provide sufficient enrollment and better reflect the patient population.

\begin{figure}[ht]
	\centering
	\caption{\glsfmtshort{best} Trial Schematic}
	\glsadd{best}
	\label{fig:best-schematic}
	\includegraphics[width=\linewidth]{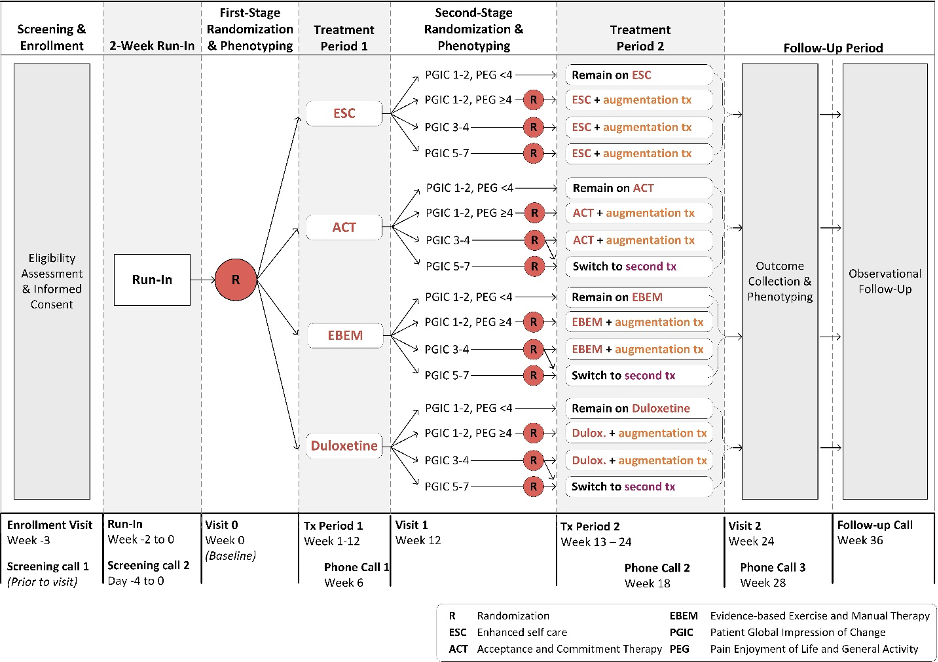}
\end{figure}

A detailed schematic of the \gls{best} trial's \gls{smart} design is available in Figure~\ref{fig:best-schematic}. In the following subsections, we highlight key features of the \gls{best} trial's \gls{smart} design and  discuss how these elements, though not unique to \gls{best}, address specific clinical and statistical questions. 

\subsection{Active Treatments Only}\label{design:active}

The primary objective of this trial was to maximize treatment effectiveness rather than confirm treatment efficacy of any individual treatment or directly compare them. The \gls{best} trial was designed to address this objective and mirror real-world clinical decision-making by exclusively studying active treatments without a placebo arm, as each selected study treatment has demonstrated effectiveness over placebo \citep{mauck2022evidence}, which is not prescribed in routine care. Consequently, interim analyses for futility or efficacy were not planned or conducted. Since the study evaluates only active treatments currently used in clinical practice, the average study outcome approximates the expected improvement under standard of care when treatment is limited to these interventions. 

\subsection{Responder Categories}\label{design:resp}

The \gls{best} trial's \gls{smart} design consisted of two 12-week treatment periods with no intervening washout, followed by a 12-week follow-up period. Participants were randomly assigned to one of four study interventions for stage one. Treatment response was assessed at the 12-week visit (Visit 1) using two validated measures: the \gls{peg} \citep{krebs2009Development} and the \gls{pgic}, a patient-reported outcome with ordinal values from 1 to 7 \citep{dworkin2005Core, farrar2001Clinical} with higher values indicating more pain. Rather than using a dichotomous response/non-response, four response categories and corresponding treatment options for stage two were defined based on the 12-week \gls{peg} and \gls{pgic}:
\begin{itemize}[nosep]
\item \gls{pgic} 1-2 and \gls{peg} $<$4: maintained first-stage treatment
\item \gls{pgic} 1-2 and \gls{peg} $\geq$4: maintained first-stage treatment and received a randomly assigned augmentation treatment
\item \gls{pgic} 3-4: randomized to either receive a randomly assigned augmentation treatment or switch to a randomly assigned new treatment
\item \gls{pgic} 5-7: switched to a randomly assigned new treatment
\end{itemize}

Participants randomized to \gls{esc} in stage one with a 12-week \gls{pgic} of 3-7 always received a new randomly assigned treatment and were considered as having augmented rather than switched their initial treatment because \gls{esc} taught self-care skills and practices that could not be unlearned or fully discontinued. Additionally, participants who discontinued their assigned first-stage treatment early were always switched to a new treatment for stage two.

A committee of subject-matter experts designed these categories to reflect clinical decision-making for continuing, switching, or augmenting treatments. Response status and first-stage treatment determined feasible second-stage interventions: best responders continued their initial treatment per clinical practice, while others were randomized to either augment or switch treatments. Although second-stage treatment assignment depended on first-stage response, this design was not response-adaptive \citep{hu2006Theory}, as randomization probabilities remained fixed rather than updating based on accumulated outcomes. We detail the randomization algorithm in the \nameref{sec:randomization} section.

\subsection{Study Population and Treatment Exclusions}\label{design:contra}

The \gls{best} trial permitted study participants to be excluded/ineligible from being randomized to one, but no more, study interventions. This decision was motivated by the number of distinct treatment modalities in the study combined with the expectations that in the \gls{clbp} patient population treatment exclusions would not be rare and participants would likely have been undergoing active treatment at the time of enrollment. Widening inclusion criteria involves balancing enhanced generalizability against potential reductions in internal validity and statistical power from increased population heterogeneity. A high frequency of contraindications could lead to treatment assignment imbalance and differential covariate distributions across treatment arms. To address this challenge while maintaining generalizability, we allow participants to be contraindicated for at most one treatment. Allowing contraindications requires attention to potential treatment and covariates imbalances. At the extreme, balance becomes impossible in a $\maxarmindex$-armed trial if the prevalence of the contraindication is greater than $(\maxarmindex - 1)/\maxarmindex$. We elaborate on how we minimized imbalances in the \nameref{sec:randomization} section.

\section{Planned Primary Analysis}\label{prim}

\subsection{Modified Intention-to-Treat Population}\label{prim:mitt}

The primary analysis will include all participants who received any component of the treatment to which they were randomized in both stage one and stage two, referred to as the \gls{mitt} population. The requirements for inclusion in the \gls{mitt} population are presented in Table~\ref{tab:mitt}. The definition of ``receiving any treatment'' is treatment-specific and, in stage two, depends on whether the treatment is being continued from stage one.  Participants assigned to augment their initial treatment in stage two must meet the requirements for both the new treatment they are starting and the one they are maintaining. 

\ifbool{isarxiv}{ 
    \begin{table}[ht]
    \centering
    \setlength{\tabcolsep}{4pt} %
    \renewcommand{\arraystretch}{0.95} %
    \begin{tabularx}{.85\textwidth}{@{}llX@{}} %
    \toprule
    Intervention & Timing & Requirements                                                                   \\ \midrule
    ACT          & Initiation & Attend at least one session                                            \\
    EBEM         & Initiation & Attend at least one session                                   \\
    ESC          & Initiation & Attend initial treatment visit and set up an online account                    \\
    Duloxetine   & Initiation  & Any evidence of having taken at least one pill  (includes
    self-report at any follow-up and returning fewer pills than provided). \\ 
    ACT          & Maintenance  & No additional requirements \\
    EBEM         & Maintenance     & Attend at least one of the continuing sessions  \\
    ESC          & Maintenance     & No additional requirements                       \\
    Duloxetine   & Maintenance    & Any evidence of having taken at least one pill       \\ \bottomrule
    \end{tabularx}
    \caption{Requirements for \gls{mitt} Population}
    \label{tab:mitt}
    \end{table}
}{ 
    \begin{table}[ht]
    \centering
    \small %
    \setlength{\tabcolsep}{4pt} %
    \renewcommand{\arraystretch}{0.95} %
    \begin{tabularx}{\textwidth}{@{}llX@{}} %
    \toprule
    Intervention & Timing & Requirements                                                                   \\ \midrule
    ACT          & Initiation & Attend at least one session                                            \\
    EBEM         & Initiation & Attend at least one session                                   \\
    ESC          & Initiation & Attend initial treatment visit and set up an online account                    \\
    Duloxetine   & Initiation  & Any evidence of having taken at least one pill.\footnote{This includes
    self-report at any of the 7-day, 6-week, or 12-week follow-ups and returning a number of pills less than initially provided.} \\ 
    ACT          & Maintenance  & No additional requirements \\
    EBEM         & Maintenance     & Attend at least one of the continuing sessions  \\
    ESC          & Maintenance     & No additional requirements                                                     \\
    Duloxetine   & Maintenance    & Any evidence of having taken at least one pill       \\ \bottomrule
    \end{tabularx}
    \caption{Requirements for \gls{mitt} Population}
    \label{tab:mitt}
    \end{table}
}

Participants who meet the requirements for both stages will remain in the \gls{mitt} population, even if they are missing the primary endpoint of the 24-week \gls{peg}. We discuss how we plan to account for missing data in the \nameref{prim:miss} section.
\subsection{DTR Estimation}\label{prim:dtr}

We will employ Q-learning \citep{watkins1992Qlearning, schulte2014Alearning} with functional estimation to estimate the \gls{odtr} for maximizing the average improvement in the 24-week \gls{peg} relative to baseline by incorporating patients features (including the interim response to the initial treatment) for the primary analysis. Q-learning is a general framework for reinforcement learning where the reinforcement learning
problem of solving a multi-stage decision problem is reduced to a series of single-stage regression problems using backwards induction. The Q-learning framework is agnostic to the choice of the algorithm or model used to estimate the regression functions
\citep{tsiatis2019Dynamic,schulte2014Alearning,chakraborty2014Dynamic}. 

The comprehensive Q-functions will be estimated using an ensemble of flexible machine-learning and statistical methods. Explainable
artificial intelligence  methods and variable importance metrics will be used to investigate these estimated models and \glspl{dtr}. We will estimate a comprehensive and explainable version and a simplified clinically-implementable, interpretable version of the \gls{odtr}. While terminology is evolving, we follow the convention of using ``explainable'' for methods that use a complex model for prediction or treatment assignment and a separate additional method to provide a simplified explanation of how the complex model makes predictions; we use “interpretable” for methods where the model used for prediction or decision making can be understood without additional methods to simplify the model.

The comprehensive ensemble approach will employ a regression ensemble—a meta-regression model that combines predictions from multiple constituent models using the Super Learner framework \citep{wolpert1992stacked,mienye2022survey}. This framework optimally weights predictions from various models through cross-validation and is theoretically guaranteed to asymptotically perform at least as well as the best individual model. The set of modeling methods included in the ensemble will be selected based on resource and software availability, without data-dependent inclusion or exclusion criteria, and will include parametric and nonparametric approaches. 

The interpretable approach will incorporate both Q-learning with interpretable models \citep{liu2022fasterrisk,hastie2009additive} for estimating Q-functions and direct methods for constructing interpretable \glspl{dtr}, developed in collaboration with statisticians and clinicians. Q-learning involves modeling the response for each intervention separately, whereas direct methods focus solely on identifying features that influence optimal treatment selection, resulting in more parsimonious models. By excluding features that are unnecessary for decision-making, direct methods may produce \glspl{dtr} that are more practical for clinical implementation, at the cost of losing detailed information about treatment-specific response predictions.

The estimated ensemble and interpretable \glspl{odtr} will both contribute valuable insights into \gls{clbp} treatment. The ensemble \gls{odtr} will be used to estimate the potential value of the \gls{odtr} for the study interventions and measurements, i.e. estimate the best that could be expected in terms of improving the 24-week \gls{peg} relative to baseline by assigning the study interventions using the measurements included in the primary analysis. The ensemble Q-functions will additionally be used to identify promising patient features for predicting treatment response. The interpretable \gls{odtr} could be used for clinical decision making if the trial results are promising, and we will estimate the potential value of the estimated interpretable \gls{odtr}. Here we are estimating the value of the estimated interpretable \gls{odtr} based on the data collected by the \gls{best} trial; in contrast, the ensemble-estimated \gls{odtr} is used to estimate the value of the the rule we would devise with perfect information rather than its own value if it were implemented.

We will report the estimated difference and corresponding 95\% \gls{ci} for the expected improvement in 24-week \gls{peg} relative to baseline between assigning treatments according to the \gls{odtr} and assigning treatments as implemented in the study (i.e., using randomization and the study's treatment assignment rules). Additionally, we will compare the values of the ensemble and interpretable models as a diagnostic check to assess potential trade-offs in predictive performance when using a simplified model. This comparison provides insight into the information that may be lost by prioritizing interpretability.

\subsection{Methods for Missing Data}\label{prim:miss}
We implemented comprehensive retention strategies to minimize missing data. These include offering incentives, creating patient-informed study materials, and reducing participant burden. To further support retention, we maintained continuous engagement through study websites, regular communication with research staff, and ongoing patient engagement initiatives. Efforts were made to follow all participants until study completion, utilizing alternative data collection methods (e.g., telephone or online) when in-person visits were not feasible. Despite these retention strategies, some amount of missing data was inevitable.

We will address missing data under the \gls{mar} assumption, anticipating that study attrition will account for most missing data and likely produce a nearly monotone missing data pattern. For the primary analysis, we will utilize \gls{mi} to address missing covariate data combined with \gls{ipw} to address missing outcome data. Multiple methods have been developed for handling missing data in both single-stage trials estimating \glspl{dtr} and multi-stage \gls{smart} designs, including approaches for missing covariates, treatments, outcomes, and their combinations \citep{shortreed2014multiple,nahum2017smart,lu2023constructing,shen2023estimation}. For \gls{mi}, we will characterize missing data based on their time points and use separate imputation models conditioned on observed information. Weighting rather than imputation will be used to address missing outcome in the primary analysis. However, the outcome will be imputed as a contingency if the weighting approach would result in a substantial number of extreme weights as determined by model diagnostics.

\section{Power and Sample Size}

Recent systematic reviews on the uptake of \glspl{smart} and
\gls{dtr} estimation in geriatric medicine and in oncology have shown that the
primary analysis for a \gls{smart} still rarely involves \gls{dtr} estimation
even when the primary objective is framed in terms of estimating a treatment rule or
mapping patients to treatment (i.e. the primary objective is \gls{dtr} estimation)
\citep{kahkoska2023Individualized, lorenzoni2023Use}. Instead, methods more
aligned with traditional inferential questions are often reported though they
may not be informative about the design's potential to answer the
\gls{dtr}-related questions. Some common approaches are:

\begin{enumerate}
	\item Comparative effectiveness of the initial treatments in terms of their average treatment effects. This can be done using only
	      the first stage data and outcomes, or by using the outcome after all stages of
	      treatment and averaging over later treatments.
	\item Comparing the efficacy of treatment options for nonresponders to the initial treatment.
	\item Comparing two or more embedded \glspl{dtr} in the study.
	      An embedded \gls{dtr} is a \gls{dtr} that involves only paths that were directly
	      assigned in the \gls{smart}.
\end{enumerate}

None of the three common approaches currently seen in the applied literature
using \glspl{smart} are appropriate to achieve the study's primary objective of
\gls{dtr} estimation involving covariates. The first is not informative about
covariates, and the \gls{best} interventions are expected to only have small
differences in terms of average treatment effects in the population as a whole. The second incorporates patient features
indirectly through their effect on responder status, and is not of primary
scientific interest to the BEST trial's clinical team due to the existing evidence for the benefit of the selected treatments.
Additionally, the first two approaches do not make use of all of the trial data. Embedded \gls{dtr}-based approaches may use covariates in constructing the embedded regimes, but the covariates must be
known ahead of time in order to do so.

The dominance of these approaches likely stems from a shortage of methods for sample size determination in \glspl{smart} where
the primary objective involves covariates. In addition to the scarcity
of methods, the lack of published interaction effect estimates for treatments
and covariates complicates sample size determination in this setting. For the
\gls{best} trial, sample size was based on simulation studies under various potential data generating models. 
\subsection{Sample Size for DTR Estimation}
We based our sample size simulation on the value function of the \gls{dtr} estimated at the end of a trial as a summary measure of the \gls{dtr}'s quality. Specifically, we compared the expected value of the estimated \gls{dtr} based on $\maxobsindex$ participants, $\val\left( \widehat{\pol}_{\maxobsindex} \right)$ 
to the value of the \gls{odtr} $\val\left( \optpol
	\right)$. The value of an estimated \gls{dtr} $\polhatnx$ quantifies the expected response by the potential outcomes 
$\pout$ under the treatment decisions made according to 
$\polhatn$:
\begin{equation}\label{eqn:val-fn}
    \val\left( \polhatnx \right) = \E_{\covarrv}[\pout(\polhatnx) \mid \polhatn]
\end{equation} 
As an expectation, the value function naturally accounts for differences in the magnitude and prevalence of heterogeneous treatment effects. 

We defined power in terms of the probability that the value of the estimated \gls{dtr} is within a set tolerance of the value of the true \gls{odtr} under a given data generating model. We used an acceptable tolerance for how close to optimal the estimated \gls{dtr} could be in terms of the percentage of optimal value,
$\val\left( \polhat_{\maxobsindex} \right)/\val\left( \optpol \right)$; the difference between the two values could also be used.
For the ratio measure, covariates should be centered so that main effects do not
contribute to the value of the \glspl{dtr}, $\val\left( \widehat{\pol}_{\maxobsindex}
	\right)$ should be non-negative, and $\val\left( \optpol \right)$ strictly positive. 

This power metric evaluates how close the estimated \gls{dtr} is to the \gls{odtr}. It provides guarantees about the design and estimation
procedures so that we can be reasonably confident that the \gls{dtr} estimated
at the end of the trial is close to optimal even though further confirmatory
trials may be needed to prove it. An appropriate hypothesis test of the value
function should be used if the trial's primary objective is confirmatory.

\subsection{Sample Size and Simulation Overview}\label{best:power-sim}

The trial aimed to enroll approximately 820 participants to obtain an estimated 740 randomized participants and $\maxobsindex = 630$ protocol completers. This assumed a 10\% loss related to post-run-in eligibility and 15\% attrition over the trial period. Under this assumption, a sample size of 630 completers would yield 81\% probability (Monte Carlo standard error$\approx0.01$) of estimating a \gls{dtr} within 90\% of the optimal regime based on the study’s primary endpoint, the \gls{peg} measured at 24 weeks.

Simulation studies were conducted to estimate the required sample size in terms of the number of completers $\maxobsindex$, and the sample size was found using grid search over the range of practically feasible sample sizes. We evaluated sample sizes in terms of the expected value of the estimated \gls{dtr} based on a trial of that sample size. This performance metric quantified how close the estimated optimal \gls{dtr} was to the true optimal regime from the data generating model. This metric required identifying features for predicting treatment response, and it had the advantage of implicitly weighing estimation error by a feature's importance to identifying the optimal treatment(s). This metric bears a family resemblance to the ``probably approximately correct'' learning framework\citep{valiant1984Theory} that could be investigated in future work. The value of the estimated \gls{dtr} for each replicate was approximated using Monte Carlo methods. 

The average treatment effect sizes for the data generating models were based on an internal review conducted by the experts on \gls{best}'s Interventions Working Group and a systematic review by 
\ifbool{isarxiv}{\cite{mcdonagh2020Nonopioid}}{McDonagh et al. 2020\cite{mcdonagh2020Nonopioid}}
.
They were 0.1 for \gls{esc}, 0.25 for ACT, 0.3 for Duloxetine, and 0.4 for EBEM. The interaction effect sizes were constrained to be less than half the size of the main effects so that the average treatment effects would remain unchanged. Simulation scenarios included sparse and dense data models where there were relatively few markers with greater effects as well as many markers with small effects, respectively. We report the sparse simulation results here as it was the most challenging scenario among those tested. 

The optimal \glspl{dtr} in the simulation were estimated using Q-learning with the Lasso regression \citep{tibshirani1996Regression}. The value of the estimated \gls{dtr} was approximated by Monte Carlo methods, and then the ratio of this value to the value of the \gls{odtr} was calculated. We summarized the performance of a design by the probability that the ratio of the value from a \gls{dtr} estimated from a trial with the given sample size of
$\maxobsindex$ relative to the value of the \gls{odtr} was greater than or equal to a fixed performance cutoff $\perfcutoff$, as shown in Equation~\ref{eq:best-opt-power}.

\begin{equation}\label{eq:best-opt-power}
	\prob\left( \val\left( \polhat_{\maxobsindex} \right) \div \val\left(
		\optpol \right) \geq \perfcutoff \right)
\end{equation}

The \gls{best} trial utilized a performance cut-off $\perfcutoff$ of 0.9. That is, for a simulated trial to be considered a success, the value of the average outcome from the estimated \gls{dtr} on that simulation replicate's data must be within 90\% of the \gls{odtr}'s average outcome value. 
Note there is no analog to Type-I error in this setting because a
null scenario would imply that all treatments are equally ineffective, no
patient features matter, and in this case any \gls{dtr} is trivially optimal. The simulations were conducted using \textrm{R} version 4.0.4 \citep{rcoreteam2021Language}, and the code is available on Github\footnote{\url{https://github.com/UNC-CSCC/bacpac_collaborativeSMARTsims}}.

\subsection{Simulation Details}

The outcome after each stage, $\resprv_{\timeindex}$, $j=1,2$, was defined as the
difference in \gls{peg} between the follow-up visit at the end of the stage and
the baseline \gls{peg}. The outcome of interest, $\resprv$, is
the difference in \gls{peg} at 24 weeks relative to baseline, $\resprv =
	\resprv_{2}$. The simulation outcomes are on the standardized effect
size scale instead of the 0-10 scale.

Let $\armrv_{\timeindex,\armindex}$ denote the indicator $\indfun \left(
	\text{Intervention}\ \armindex \text{ given for stage }\timeindex \right)$
which is one whenever $\text{intervention}\ \armindex \text{ is given for stage
	}\timeindex$ and zero otherwise. For the outcome after each stage,
$\resprv_{\timeindex}$, denote the conditional expectation of
$\resprv_{\timeindex}$ given the history of previous treatments and covariates, $H_j$, and intervention by

\begin{align}
	Q_{1}\left( \histrv_{1},\armrv_{1} \right) & = 0.1\armrv_{1,0} + 0.25\armrv_{1,1}
	+ 0.3\armrv_{1,2} + 0.4\armrv_{1,3}                                                                                       \label{eq:q1} \\ &+ 0.3\covarrv_{1}\armrv_{1,3} +
	0.25\covarrv_{2} - 0.25\covarrv_{3} \nonumber                                                                                           \\
	Q_{2}\left( \histrv_{2},\armrv_{2} \right) & = 0.1\armrv_{2,0} + 0.25\armrv_{2,1}
	+ 0.3\armrv_{2,2} +
	0.4\armrv_{2,3}                                                                                       \label{eq:q2}                    \\&+ 0.3\covarrv_{1}\armrv_{2,3} -
	0.1\armrv_{2,1}\armrv_{2,2} - 1\armrv_{2,1}\armrv_{2,3} \nonumber                                                                      \\
	                                           & - 0.1\armrv_{2,2}\armrv_{2,3} + 0.25\covarrv_{2} - 0.25\covarrv_{3} \nonumber \\
	\resprv_{1} & = Q_{1}\left( \histrv_{1},\armrv_{1} \right) + \err_{1},\ \err_{1}
	\sim N(0,1) \nonumber                                  \\
	\resprv_{2} & = Q_{2}\left( \histrv_{2},\armrv_{2} \right) + \err_{2},\ \err_{2} 
	\sim N(0,1) \nonumber 
\end{align}

\begin{figure}[ht]
	\centering
	\caption{BEST Trial Sample Size Simulation Results}

	\includegraphics[width=.7\textwidth]{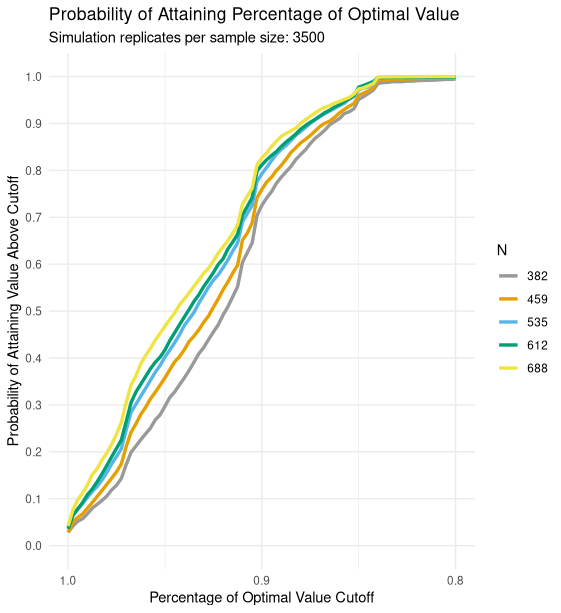}
	
	\label{fig:best-sample-size}
\end{figure}

We simulated a total of ten covariates, one of which was associated with the outcome through a
treatment interaction, and two prognostic covariates that influence the outcome
but were not associated with response to treatment. The other covariates were considered nuissance.
Five covariates were generated independently with $\covarrv_{l} \sim \text{Bern}(p_{l})$ for $l = 1,
	..., 5$ with $p_{l}$ ranging from 0.5 to 0.9 and an additional
five were normally distributed $N(0,\ .5)$ covariates.
The $\covarrv_{l}$ were coded to take values $\{ - 1, 1\}$ instead of $\{ 0, 1\}$.
The $Q$-functions in equations \ref{eq:q1} and \ref{eq:q2} were
estimated using mis-specified models that included interaction terms for every
treatment-by-covariate pair, and the models were fit using Lasso regression. The results of the simulation study varying the number of participants with
outcome data for both stages is shown in
Figure~\ref{fig:best-sample-size} with 3,500 simulation replicates performed
for each $\maxobsindex$. An additional simulation with 5,000
replicates was conducted with 630 completers to confirm the estimated power was at least two Monte Carlo standard deviations above the target of 80\%.

\section{Randomization}\label{sec:randomization}

\subsection{Randomization Design}
The \gls{best} trial used a minimization algorithm \citep{pocock1975Sequential, taves1974Minimization} with a biased coin flip for randomization in both stages (R+M). Minimization offers an alternative to stratified randomization for reducing imbalances across treatments and covariates by assigning a higher probability to the treatment(s) that would most reduce a prespecified imbalance measure with the other treatments having an equal, lower assignment probability. In all cases where minimization was used, ties for the largest imbalance reduction were broken equally at random before biased coin randomization. We used the R+M procedure in Stage 1 to assign individuals to their initial treatment, in Stage 2 to assign participants with Visit 1 PGIC=3,4 to augment or switch, and in Stage 2 to assign treatment when it was not algorithmically determined (i.e. in the case of maintaining). 

We modified the standard minimization method to allow subjects who are ineligible to receive certain study interventions as detailed in Algorithm~\ref{alg:minim-contra}. In \gls{best}, patients could be excluded from one of the study treatments at baseline (see the \nameref{design:contra} section), but the method could be applied in studies with more restrictions as indicated by the use of a feasible arm set in Algorithm~\ref{alg:minim-contra}. The adapted algorithm calculated the imbalance measure for each participant $\obsindex$ at randomization. The algorithm included all treatment options when computing the imbalance scores, but after calculation the biased-coin randomization only included the feasible treatments. Minimization algorithms require specifying an imbalance measure $\minimfunc$ that is non-negative and where larger values indicate greater imbalance. The \gls{best} trial used the marginal discrepancy over the four binary factors for the measure of imbalance. Minimization-based methods can accommodate continuous covariates by selecting an imbalance metric appropriate for continuous covariates such as the Euclidean norm of the variance matrix.  

\subsection{Implementation}
In Stage 1 to randomize to initial treatment, \Gls{best} used four binary factors measured at baseline in the minimization: depressive or anxiety symptoms (yes / no), duration of pain in years ($<$ 5 / $\geq$ 5), and current use of opioids (yes / no) were selected because they were thought to be highly predictive of the change in \gls{peg} but not predictive of treatment response for the four study intervention modalities. The fourth factor was the participant's willingness to participate in additional optional phenotyping (yes / no) to ensure balance of optional phenotyping data collection across arms. In Stage 2, the same four binary factors collected at Visit 1 were used in the minimization, except in treatment minimization for participants with Visit 1 PGIC=3,4. The minimization procedure for Stage 2 treatment for these individuals also included whether or not they had been randomized to augment or switch. The randomization probability $\minimprob$ favoring assignments that minimize imbalances and weights $\minimweight$ and $\minimarmweight$ for balancing factors and treatment arms to adjust their relative importance. In Stage 1, by default the $\minimprob$ is equal to $2/3$, and all weights are equal to one for $\minimweight$ and $\minimarmweight$. In Stage 2,  the $\minimprob$ is equal to $2/3$, and duration of pain and consent to optional phenotyping had a weight of 2 and all other factors had a weight of 1 for $\minimweight$ and $\minimarmweight$ to accommodate balance for more important factors in smaller sample sizes after stratifying on Stage 1 treatment and treatment response. Across all stages, site was not used as part of the randomization because study sites may have constraints that limited their capacity to certain interventions. We chose to prioritize study-wide balance instead of trying to enforce site-level balance.

\begin{algorithm}[ht]
\SetKwInOut{Input}{Input}
\SetKwInOut{Output}{Output}
\SetKwInOut{Optional}{Optional}
	\caption{Minimization with Treatment Exclusions Algorithm Step for a Single Treatment Assignment}\label{alg:minim-contra}
        \Input{Assignment and covariate history $H_{\obsindex} = \{(x_i, a_i, )\}_{i = 1}^{t-1}$, Set of all study interventions $\mathcal{A}$,\newline Imbalance measure $g()$, \newline Biased coin favored probability $\minimprob$}
        \Optional{Measure-specific parameters $\omega$}
        \Output{Treatment Assignment for Participant $\obsindex$, $a_{\obsindex} \in \mathcal{A}_{\obsindex}$ }
	\DontPrintSemicolon
	Observe new participant $\obsindex$'s covariates $x_{\obsindex}$ and set of treatments they are eligible for $\mathcal{A}_{\obsindex} \subseteq \mathcal{A}$ \;
      
\For{every arm $k \in \mathcal{A}$}{
    Define the hypothetical design matrix $X_{t,k}^{\star}$ if participant $t$ were assigned to arm $k$\;
    Calculate the resulting imbalance score $S_{k}$ for arm $k$: 
    $S_{k} = \minimfunc(X_{t,k}^{\star}, \minimweight)$\;
}
		Find the treatment arm with the minimum imbalance score from the set of
		feasible arms: $\armindex^* = \argmin_{\mathcal{A}_{\obsindex}} S_{\armindex}$\;
		Randomly assign the participant to treatment arm $\armindex^*$ with probability
	$\minimprob$ or to another feasible arm each with probability
	$(1-\minimprob)/(|\mathcal{A}_t - 1|)$\;
\end{algorithm}

The choice of imbalance measure for the \gls{best} trial was the marginal discrepancy. The formula for calculating the marginal discrepancy when the covariates are binary with zero-one coding is given below in equation~\ref{eqn:marginal-disc}. 

\begin{equation}\label{eqn:marginal-disc}
	S_{\armindex} = \max_{\armindexprime \neq k} \sum_{\minimdim}
	\minimweight_{\minimdim}| (n_{\armindex,
			\minimdim} + \covarrv_{\minimdim}) - n_{\armindexprime, \minimdim}|
\end{equation}

where $n_{\armindex, \minimdim}$ is the number of participants with value one for covariate $\covarrv_{\minimdim}$ assigned to treatment $\armindex$, $\minimweight_{\minimdim}$ is the relative weight assigned to balancing factor $\minimdim$.

\subsection{Imbalance Simulation}\label{best:imbalance-sim}

A simulation study was conducted to evaluate the effectiveness of the modified minimization method with contraindications. Each simulated replication
involved a sample size of $\maxobsindex = 600$ participants while the number of baseline
covariates was varied to assess how many covariates could feasibly be included in the minimization algorithm. The marginal distributions of the baseline covariates were all Bernoulli with probability $p=.5$ and the baseline covariates were correlated within site to reflect the
potential for regional differences in covariate distributions. Eight sites were used in the simulation with an intra-class correlation of .05; the simulation study was conducted before the final determination of 12 sites. For the algorithm, the probability of assigning the favored treatment $\minimprob$ was equal to $2/3$, and all weights were equal to one.

Let $n_{\armindex_{ij}}$ denote the number of patients that are on treatment $\armindex$ for the
$j$-th level of prognostic factor $i$. The number of patients on a treatment can be recovered by summing
over the  $n_{\armindex} = \sum_{i = 1}^I \sum_{j = 1}^{J_i}n_{\armindex_{ij}}$. Let $n_{\armindex_{s}}$ denote the number of
patients that are on treatment $\armindex$ at site $s$. We investigated covariate imbalances at three levels:

\begin{itemize}
	\item Study-wide treatment imbalance: $$\max_{\armindex, \armindexprime \in [\maxarmindex]} |n_{\armindex} - n_{\armindexprime}|$$
	\item (Covariate) Factor-level imbalance: $$\max_{\armindex, \armindexprime \in [\maxarmindex]}|n_{\armindex_{ij}} - n_{\armindexprime_{ij}}|$$
	\item Site-treatment-level imbalance: $$\max_{s} \max_{\armindex, \armindexprime \in [\maxarmindex]}|n_{\armindex_s} - n_{\armindexprime_s}|$$
\end{itemize}

The study-wide imbalance is the difference between the arm with the most participants and the arm with the fewest. The factor-level imbalance looks at the maximal imbalance across each covariate included in the minimization; for example, the difference between the number of participants with depressive or anxiety symptoms assigned to \gls{act} versus \gls{ebem}. 

\begin{figure}[ht]
	\centering
	\caption{Boxplots of Covariate Imbalance Using Minimization with Contraindications}
	\label{fig:best-imbalance}
	\includegraphics[width=.8\linewidth]{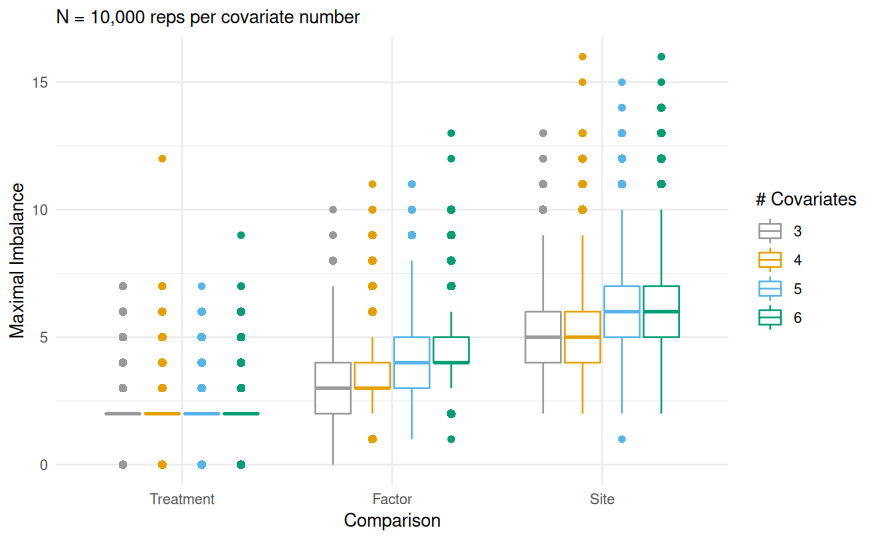}
\end{figure}

The results from 10,000 simulation replications per covariate setting are shown
in Figure~\ref{fig:best-imbalance}. For all numbers of baseline covariates
tested, the maximal imbalance was quite small. For the overall treatment
imbalance, the distribution of the maximal imbalance was concentrated around
two with the 1st and 3rd quartiles and median all equal to two. Across all
simulated values of baseline covariates included in the minimization algorithm there was only a single study replicate with a maximal treatment imbalance of 10 or more out of 40,000 replications. In other words, more than 99.9\% of simulated studies had a difference of fewer than 10 people between the study arm with the most participants and the arm with the fewest participants in the first stage. 

Results for the factor-level imbalance and site-treatment-level imbalance are
mildly less dramatic and still well controlled. For the \gls{best} trial's
number of baseline covariates, four, the median maximal imbalance
at the factor-level was three and the median site-level treatment imbalance was
five indicating indicating a high level of balance across treatment arms.

\section{Discussion}\label{sec:discussion}

The \gls{best} trial advances precision medicine in \gls{clbp} through innovative design choices that address unique challenges in chronic pain research. Using a \gls{smart} design, the study enables the exploration of treatment strategies tailored to individual patient characteristics and their responses to treatment. These results could directly inform clinical practice through evidence-based treatment recommendations. Findings from the trial will advance \gls{clbp} research, as the estimated optimal treatment rule could serve as the basis for future confirmatory trials assessing its efficacy as a treatment guideline. Additionally, the \gls{best} trial could enhance our understanding of pain mechanisms and treatment pathways by identifying patient subgroups with favorable responses to specific treatments or features that predict treatment response.

The proposed minimization randomization method allowed us to accommodate participants unable to receive one of the study interventions while protecting against imbalances between arms in important prognostic features. The study population better represents the broader clinical population with \gls{clbp} as a result. While the \gls{best} trial aims for broad applicability through pragmatic design choices, such as allowing single contraindications and adopting inclusive eligibility criteria, individuals with the most severe \gls{clbp} may be underrepresented due to the intensive participation requirements and or by having multiple treatment exclusions. A statistical limitation is that the imbalance simulation was conducted before the number of sites was finalized and uses 8 instead of 12 sites. 

Our simulation study allowed us to estimate the sample size in a setting where the optimal treatment(s) depended on patient features. While the past two decades of research have firmly established the theoretical foundations of estimating \glspl{dtr}, many applied and methodological statistical questions remain. Sample size methods for trials aiming to estimate a \gls{dtr} using patient features are notably underdeveloped compared to methods comparing predefined rules or embedded regimes. A key practical challenge with simulations in this setting is that published estimates don't exist for many simulation parameters e.g. carryover effects or augmentation treatment effects. 

The implications of focusing on the value function versus some alternative metric warrants further research. The value function is attractive because it is patient-centric, closely tied to clinical decision making, and directly based on the expected improvement in the primary outcome. However, its relationship with performance on other scientific tasks---such as ranking promising predictors for future study---is unclear. The value function doesn't directly penalize model size or incorporate cost information unless the primary outcome is replaced with some composite outcome incorporating these additional features.

The relationship between the scenario settings such as the number of features that affect treatment response, the sample size, and the performance tolerance and the difficulty in estimating the optimal \gls{dtr} requires further study. Sparse scenarios, where there is a small number of important features, were more challenging in our testing because the large performance penalty for missing a feature outweighed having stronger signals. In dense scenarios, where a large number of features contribute smaller effects, correctly identifying specific features may be more challenging but omitting one is less impactful. It's plausible however that dense scenarios could become increasingly challenging with more stringent success tolerances. 

The \gls{best} trial represents a landmark in \gls{clbp} research, setting new standards for statistical rigor and innovation in clinical trial design and analysis. It will generate high-quality data for clinical decision-making through multiple treatment stages, comparison of active treatments within a single study, and a thorough collection of patient-reported outcomes alongside biomarker and phenotypic data. Integration with \gls{bacpac} consortium-wide data collection protocols ensures compatibility for future research synthesis, addressing a critical gap in the field. This comprehensive approach to data collection and discovery will serve as a robust foundation for future research while yielding actionable insights to improve patient care.\label{LastPage}

\begin{dci}
\noindent J.S., K.M.K., B.Z., A.B., T.S.C., A.I., S.J.B, B.R., J.P.Z., and M.R.K. have no conflicts of interest to disclose.

\noindent M.C.M receives funding from the National Institutes of Health for research (U24AR076730, 1R21MD016467-01, 1K23AT011389-01A1).

\noindent K.J.A. Prior research funding from NIH, PCORI, Merck, and Bayer.

\noindent D.J.C. performs consulting for Merck, Eli Lilly and Company, Virios Therapeutics, and Tonix Pharmaceuticals.

\noindent N.L.B.F. has in the last 36 months or is currently funded by NIH, Amgen Inc., and Novartis. I have received speaking honoraria from Outset (Startup Summer LLC) and Saint Louis University. 

\noindent C.M.G. receives research funding to her institution from the National Institutes of Health.

\noindent M.P. is employed by GlaxoSmithKline, LLC and owns GlaxoSmithKline, LLC stock.

\noindent G.A.S. receives funding for research from the National Institutes of Health (U19 AR076725).

\noindent A.D.W. Consultant, Vertex Pharmaceuticals and SEIKAGAKU North America.
\end{dci}

\begin{funding}
This research was supported by the National Institutes of Health through the NIH HEAL Initiative under award number U24AR076730 and partially supported by the National Center for Advancing Translational Sciences (NCATS), National Institutes of Health, through Grant Award Number UM1TR004406. The Back Pain Consortium Research Program is administered by the National Institute of Arthritis and Musculoskeletal and Skin Diseases. This research is solely the responsibility of the authors and does not necessarily represent the official views of the National Institutes of Health, its NIH HEAL Initiative, or the NCATS.

\end{funding}

\begin{ethics}
The protocol for the BEST trial was approved by the Back Pain Consortium Steering Committee in July
2021. Utilizing Advarra, Inc. as a single IRB for the trial, the protocol received IRB approval in November
2021 (IRB\# 21-1972). The BEST Data and Safety Monitoring Board approved the protocol in February
2022. Each institution relied on the single IRB, and each site secured regulatory approval prior to study
initiation. The study was registered on ClinicalTrials.gov (NCT05396014). 
\end{ethics}

\ifbool{isarxiv}{\bibliographystyle{unsrtnat}
}{\bibliographystyle{template/SageV}}
\bibliography{my-library}

\begin{thebibliography}{42}
\providecommand{\natexlab}[1]{#1}
\providecommand{\url}[1]{\texttt{#1}}
\expandafter\ifx\csname urlstyle\endcsname\relax
  \providecommand{\doi}[1]{doi: #1}\else
  \providecommand{\doi}{doi: \begingroup \urlstyle{rm}\Url}\fi

\bibitem[Hoy et~al.(2014)Hoy, March, Brooks, Blyth, Woolf, Bain, Williams, Smith, Vos, Barendregt, Murray, Burstein, and Buchbinder]{hoy2014Global}
Damian Hoy, Lyn March, Peter Brooks, Fiona Blyth, Anthony Woolf, Christopher Bain, Gail Williams, Emma Smith, Theo Vos, Jan Barendregt, Chris Murray, Roy Burstein, and Rachelle Buchbinder.
\newblock The global burden of low back pain: Estimates from the {{Global Burden}} of {{Disease}} 2010 study.
\newblock \emph{Annals of the Rheumatic Diseases}, 73\penalty0 (6):\penalty0 968--974, June 2014.
\newblock ISSN 0003-4967, 1468-2060.
\newblock \doi{10.1136/annrheumdis-2013-204428}.

\bibitem[Deyo et~al.(2015)Deyo, Dworkin, Amtmann, Andersson, Borenstein, Carragee, Carrino, Chou, Cook, Delitto, Goertz, Khalsa, Loeser, Mackey, Panagis, Rainville, Tosteson, Turk, Von~Korff, and Weiner]{deyo2015Report}
Richard~A. Deyo, Samuel~F. Dworkin, Dagmar Amtmann, Gunnar Andersson, David Borenstein, Eugene Carragee, John Carrino, Roger Chou, Karon Cook, Anthony Delitto, Christine Goertz, Partap Khalsa, John Loeser, Sean Mackey, James Panagis, James Rainville, Tor Tosteson, Dennis Turk, Michael Von~Korff, and Debra~K. Weiner.
\newblock Report of the {{NIH Task Force}} on research standards for chronic low back pain.
\newblock \emph{Physical Therapy}, 95\penalty0 (2):\penalty0 e1--e18, February 2015.
\newblock ISSN 1538-6724.
\newblock \doi{10.2522/ptj.2015.95.2.e1}.

\bibitem[Von~Korff and Saunders(1996)]{von1996Course}
Michael Von~Korff and Kathleen Saunders.
\newblock The course of back pain in primary care.
\newblock \emph{Spine}, 21\penalty0 (24):\penalty0 2833--2837, 1996.

\bibitem[Hirase et~al.(2021)Hirase, Hirase, Ling, Kuo, Hernandez, Giwa, and Marco]{hirase2021Duloxetine}
Takashi Hirase, Jessica Hirase, Jeremiah Ling, Peggy~H Kuo, Gilbert~A Hernandez, Kayode Giwa, and Rex Marco.
\newblock Duloxetine for the treatment of chronic low back pain: a systematic review of randomized placebo-controlled trials.
\newblock \emph{Cureus}, 13\penalty0 (5), 2021.

\bibitem[Chou et~al.(2009)Chou, Loeser, Owens, Rosenquist, Atlas, Baisden, Carragee, Grabois, Murphy, Resnick, et~al.]{chou2009Interventional}
Roger Chou, John~D Loeser, Douglas~K Owens, Richard~W Rosenquist, Steven~J Atlas, Jamie Baisden, Eugene~J Carragee, Martin Grabois, Donald~R Murphy, Daniel~K Resnick, et~al.
\newblock Interventional therapies, surgery, and interdisciplinary rehabilitation for low back pain: an evidence-based clinical practice guideline from the american pain society.
\newblock \emph{Spine}, 34\penalty0 (10):\penalty0 1066--1077, 2009.

\bibitem[Chou et~al.(2017)Chou, Deyo, Friedly, Skelly, Weimer, Fu, Dana, Kraegel, Griffin, and Grusing]{chou2017Systemic}
Roger Chou, Richard Deyo, Janna Friedly, Andrea Skelly, Melissa Weimer, Rochelle Fu, Tracy Dana, Paul Kraegel, Jessica Griffin, and Sara Grusing.
\newblock Systemic pharmacologic therapies for low back pain: a systematic review for an american college of physicians clinical practice guideline.
\newblock \emph{Annals of internal medicine}, 166\penalty0 (7):\penalty0 480--492, 2017.

\bibitem[Qaseem et~al.(2017)Qaseem, Wilt, McLean, Forciea, and of~the American College~of Physicians*]{qaseem2017Noninvasive}
Amir Qaseem, Timothy~J Wilt, Robert~M McLean, Mary~Ann Forciea, and Clinical Guidelines~Committee of~the American College~of Physicians*.
\newblock Noninvasive treatments for acute, subacute, and chronic low back pain: a clinical practice guideline from the american college of physicians.
\newblock \emph{Annals of internal medicine}, 166\penalty0 (7):\penalty0 514--530, 2017.

\bibitem[{NIH HEAL Initiative}(2018)]{nihhealinitiative2018RFAAR19027}
{NIH HEAL Initiative}.
\newblock {{RFA-AR-19-027}}: {{HEAL Initiative}}: {{Back Pain Consortium}} ({{BACPAC}}) {{Research Program Data Integration}}, {{Algorithm Development}} and {{Operations Management Center}} ({{U24 Clinical Trial Not Allowed}}).
\newblock https://grants.nih.gov/grants/guide/rfa-files/rfa-ar-19-027.html, December 2018.

\bibitem[Mauck et~al.(2023)Mauck, Lotz, Psioda, Carey, Clauw, Majumdar, Marras, Vo, Aylward, Hoffmeyer, Zheng, Ivanova, McCumber, Carson, Anstrom, Bowden, Dalton, Derr, Dufour, Fields, Fritz, Hassett, Harte, Hue, Krug, Loggia, Mageswaran, McLean, Mitchell, O'Neill, Pedoia, Quirk, Rhon, Rieke, Shah, Sowa, Spiegel, Wasan, Wey, and LaVange]{mauck2023Back}
Matthew~C Mauck, Jeffrey Lotz, Matthew~A Psioda, Timothy~S Carey, Daniel~J Clauw, Sharmila Majumdar, William~S Marras, Nam Vo, Ayleen Aylward, Anna Hoffmeyer, Patricia Zheng, Anastasia Ivanova, Micah McCumber, Christiane Carson, Kevin~J Anstrom, Anton~E Bowden, Diane Dalton, Leslie Derr, Jonathan Dufour, Aaron~J Fields, Julie Fritz, Afton~L Hassett, Steven~E Harte, Trisha~F Hue, Roland Krug, Marco~L Loggia, Prasath Mageswaran, Samuel~A McLean, Ulrike~H Mitchell, Conor O'Neill, Valentina Pedoia, David~Adam Quirk, Daniel~I Rhon, Viola Rieke, Lubdha Shah, Gwendolyn Sowa, Brennan Spiegel, Ajay~D Wasan, Hsiao-Ying~(Monica) Wey, and Lisa LaVange.
\newblock The {{Back Pain Consortium}} ({{BACPAC}}) {{Research Program}}: {{Structure}}, {{Research Priorities}}, and {{Methods}}.
\newblock \emph{Pain Medicine}, 24\penalty0 (Supplement\_1):\penalty0 S3--S12, August 2023.
\newblock ISSN 1526-4637.
\newblock \doi{10.1093/pm/pnac202}.

\bibitem[Carter and Criswell(2023)]{carter2023back}
Robert~H Carter and Lindsey~A Criswell.
\newblock The back pain consortium (bacpac) research program: Cover.
\newblock \emph{Pain Medicine}, 24\penalty0 (Supplement\_1):\penalty0 S1--S2, 2023.

\bibitem[Mauck et~al.(2025)Mauck, Barth, Bell, Brooks, Chadwick, Gunn, Hurley, Ivanova, Piva, Schneider, Bailey, Bagaason, Batorksy, Borckardt, Bowden, Carey, Castellanos, Chen, Chidgey, Dalton, Dufour, Fields, Fritz, Goolsby, Greco, Harris, Harte, Hassett, Hoffmeyer, Berkeley, Kaplan, Kidwell, Knapik, Kosorok, Kurillo, Lobo, Lotz, Mackey, Mageswaran, Majumdar, Mao, Marras, McCumber, McLean, Mehling, Mitchell, Napadow, O'Neill, Patel, Peltier, Psioda, Rowland, Rundell, Schrepf, Sperger, Vo, Wallace, Wasan, Weaver, Weber, Williams, Wilson, Zeidan, Zhao, Clauw, and Sowa]{mauck2025Design}
Matthew~C. Mauck, Kelly~S. Barth, Kevin~M. Bell, Amber~K. Brooks, Andrea~L. Chadwick, Cameron~A. Gunn, Robert~W. Hurley, Anastasia Ivanova, Sara~R. Piva, Michael~J. Schneider, Jeannie~F. Bailey, Sarah Bagaason, Anna Batorksy, Jeffrey~J. Borckardt, Anton~E. Bowden, Timothy~S. Carey, Joel Castellanos, Lucy Chen, Brooke Chidgey, Diane Dalton, Jonathan~S. Dufour, Aaron~J. Fields, Julie~M. Fritz, Rachel~West Goolsby, Carol~M. Greco, Richard~E. Harris, Steven Harte, Afton~L. Hassett, Anna Hoffmeyer, Sara~Jones Berkeley, Chelsea Kaplan, Kelley~M. Kidwell, Gregory~G. Knapik, Michael~R. Kosorok, Gregorij Kurillo, Remy Lobo, Jeffrey~C. Lotz, Sean Mackey, Prasath Mageswaran, Sharmila Majumdar, Jianren Mao, William~S. Marras, Micah McCumber, Samuel~A. McLean, Wolf Mehling, Ulrike~H. Mitchell, Vitaly~J. Napadow, Conor O'Neill, Kushang~V. Patel, Scott Peltier, Matthew Psioda, Bryce Rowland, Sean~D. Rundell, Andrew Schrepf, John Sperger, Nam Vo, Mark~S. Wallace, Ajay~D. Wasan, Tristan~E. Weaver, Kenneth~A. Weber, David~A.
  Williams, Leslie Wilson, Fadel Zeidan, Beibo Zhao, Daniel~J. Clauw, and Gwendolyn~A. Sowa.
\newblock The design and rationale of the biomarkers for evaluating spine treatments (best) trial.
\newblock \emph{Under Review}, 2025.

\bibitem[Engel(1977)]{engel1977Need}
George~L. Engel.
\newblock The {{Need}} for a {{New Medical Model}}: {{A Challenge}} for {{Biomedicine}}.
\newblock \emph{Science}, 196\penalty0 (4286):\penalty0 129--136, April 1977.
\newblock \doi{10.1126/science.847460}.

\bibitem[Gatchel et~al.(2007)Gatchel, Peng, Peters, Fuchs, and Turk]{gatchel2007Biopsychosocial}
Robert~J. Gatchel, Yuan~Bo Peng, Madelon~L. Peters, Perry~N. Fuchs, and Dennis~C. Turk.
\newblock The biopsychosocial approach to chronic pain: Scientific advances and future directions.
\newblock \emph{Psychological Bulletin}, 133\penalty0 (4):\penalty0 581--624, 2007.
\newblock ISSN 0033-2909.
\newblock \doi{10.1037/0033-2909.133.4.581}.

\bibitem[Batorsky et~al.(2023)Batorsky, Bowden, Darwin, Fields, Greco, Harris, Hue, Kakyomya, Mehling, O’Neill, et~al.]{batorsky2023back}
Anna Batorsky, Anton~E Bowden, Jessa Darwin, Aaron~J Fields, Carol~M Greco, Richard~E Harris, Trisha~F Hue, Joseph Kakyomya, Wolf Mehling, Conor O’Neill, et~al.
\newblock The back pain consortium (bacpac) research program data harmonization: Rationale for data elements and standards.
\newblock \emph{Pain Medicine}, 24\penalty0 (Supplement\_1):\penalty0 S95--S104, 2023.

\bibitem[Krebs et~al.(2009)Krebs, Lorenz, Bair, Damush, Wu, Sutherland, Asch, and Kroenke]{krebs2009Development}
Erin~E. Krebs, Karl~A. Lorenz, Matthew~J. Bair, Teresa~M. Damush, Jingwei Wu, Jason~M. Sutherland, Steven~M. Asch, and Kurt Kroenke.
\newblock Development and {{Initial Validation}} of the {{PEG}}, a {{Three-item Scale Assessing Pain Intensity}} and {{Interference}}.
\newblock \emph{Journal of General Internal Medicine}, 24\penalty0 (6):\penalty0 733--738, June 2009.
\newblock ISSN 0884-8734.
\newblock \doi{10.1007/s11606-009-0981-1}.

\bibitem[Mauck et~al.(2022)Mauck, Aylward, Barton, Birckhead, Carey, Dalton, Fields, Fritz, Hassett, Hoffmeyer, et~al.]{mauck2022evidence}
Matthew~C Mauck, Aileen~F Aylward, Chloe~E Barton, Brandon Birckhead, Timothy Carey, Diane~M Dalton, Aaron~J Fields, Julie Fritz, Afton~L Hassett, Anna Hoffmeyer, et~al.
\newblock Evidence-based interventions to treat chronic low back pain: treatment selection for a personalized medicine approach.
\newblock \emph{Pain Reports}, 7\penalty0 (5):\penalty0 e1019, 2022.

\bibitem[George et~al.(2020)George, Goertz, Hastings, and Fritz]{george2020transforming}
Steven~Z George, Christine Goertz, S~Nicole Hastings, and Julie~M Fritz.
\newblock Transforming low back pain care delivery in the united states.
\newblock \emph{Pain}, 161\penalty0 (12):\penalty0 2667--2673, 2020.

\bibitem[Wang et~al.(2012)Wang, Rotnitzky, Lin, Millikan, and Thall]{wang2012Evaluation}
Lu~Wang, Andrea Rotnitzky, Xihong Lin, Randall~E. Millikan, and Peter~F. Thall.
\newblock Evaluation of {{Viable Dynamic Treatment Regimes}} in a {{Sequentially Randomized Trial}} of {{Advanced Prostate Cancer}}.
\newblock \emph{Journal of the American Statistical Association}, 107\penalty0 (498):\penalty0 493--508, June 2012.
\newblock ISSN 0162-1459.
\newblock \doi{10.1080/01621459.2011.641416}.

\bibitem[Tsiatis et~al.(2019)Tsiatis, Davidian, Holloway, and Laber]{tsiatis2019Dynamic}
Anastasios~A. Tsiatis, Marie Davidian, Shannon~T. Holloway, and Eric~B. Laber.
\newblock \emph{Dynamic {{Treatment Regimes}}: {{Statistical Methods}} for {{Precision Medicine}}}.
\newblock {Chapman and Hall/CRC}, {New York}, December 2019.
\newblock ISBN 978-0-429-19269-2.
\newblock \doi{10.1201/9780429192692}.

\bibitem[Sperger et~al.(2020)Sperger, Freeman, Jiang, Bang, {de Marchi}, and Kosorok]{sperger2020Future}
John Sperger, Nikki L.~B. Freeman, Xiaotong Jiang, David Bang, Daniel {de Marchi}, and Michael~R. Kosorok.
\newblock The future of precision health is data-driven decision support.
\newblock \emph{Statistical Analysis and Data Mining: The ASA Data Science Journal}, 13\penalty0 (6):\penalty0 537--543, 2020.
\newblock ISSN 1932-1872.
\newblock \doi{10.1002/sam.11475}.

\bibitem[Dworkin et~al.(2005)Dworkin, Turk, Farrar, Haythornthwaite, Jensen, Katz, Kerns, Stucki, Allen, Bellamy, Carr, Chandler, Cowan, Dionne, Galer, Hertz, Jadad, Kramer, Manning, Martin, McCormick, McDermott, McGrath, Quessy, Rappaport, Robbins, Robinson, Rothman, Royal, Simon, Stauffer, Stein, Tollett, Wernicke, and Witter]{dworkin2005Core}
Robert~H. Dworkin, Dennis~C. Turk, John~T. Farrar, Jennifer~A. Haythornthwaite, Mark~P. Jensen, Nathaniel~P. Katz, Robert~D. Kerns, Gerold Stucki, Robert~R. Allen, Nicholas Bellamy, Daniel~B. Carr, Julie Chandler, Penney Cowan, Raymond Dionne, Bradley~S. Galer, Sharon Hertz, Alejandro~R. Jadad, Lynn~D. Kramer, Donald~C. Manning, Susan Martin, Cynthia~G. McCormick, Michael~P. McDermott, Patrick McGrath, Steve Quessy, Bob~A. Rappaport, Wendye Robbins, James~P. Robinson, Margaret Rothman, Mike~A. Royal, Lee Simon, Joseph~W. Stauffer, Wendy Stein, Jane Tollett, Joachim Wernicke, and James Witter.
\newblock Core outcome measures for chronic pain clinical trials: {{IMMPACT}} recommendations.
\newblock \emph{PAIN}, 113\penalty0 (1):\penalty0 9, January 2005.
\newblock ISSN 0304-3959.
\newblock \doi{10.1016/j.pain.2004.09.012}.

\bibitem[Farrar et~al.(2001)Farrar, Young, LaMoreaux, Werth, and Poole]{farrar2001Clinical}
John~T Farrar, James~P Young, Linda LaMoreaux, John~L Werth, and R.~Michael Poole.
\newblock Clinical importance of changes in chronic pain intensity measured on an 11-point numerical pain rating scale.
\newblock \emph{PAIN}, 94\penalty0 (2):\penalty0 149--158, November 2001.
\newblock ISSN 0304-3959.
\newblock \doi{10.1016/S0304-3959(01)00349-9}.

\bibitem[Hu and Rosenberger(2006)]{hu2006Theory}
Feifang Hu and William~F. Rosenberger.
\newblock \emph{The {{Theory}} of {{Response-Adaptive Randomization}} in {{Clinical Trials}}}.
\newblock {John Wiley \& Sons}, September 2006.
\newblock ISBN 978-0-470-05587-8.

\bibitem[Watkins and Dayan(1992)]{watkins1992Qlearning}
Christopher J. C.~H. Watkins and Peter Dayan.
\newblock Q-learning.
\newblock \emph{Machine Learning}, 8\penalty0 (3):\penalty0 279--292, May 1992.
\newblock ISSN 1573-0565.
\newblock \doi{10.1007/BF00992698}.

\bibitem[Schulte et~al.(2014)Schulte, Tsiatis, Laber, and Davidian]{schulte2014Alearning}
Phillip~J. Schulte, Anastasios~A. Tsiatis, Eric~B. Laber, and Marie Davidian.
\newblock Q- and {{A-learning Methods}} for {{Estimating Optimal Dynamic Treatment Regimes}}.
\newblock \emph{Statistical science : a review journal of the Institute of Mathematical Statistics}, 29\penalty0 (4):\penalty0 640--661, November 2014.
\newblock ISSN 0883-4237.
\newblock \doi{10.1214/13-STS450}.

\bibitem[Chakraborty and Murphy(2014)]{chakraborty2014Dynamic}
Bibhas Chakraborty and Susan~A. Murphy.
\newblock Dynamic {{Treatment Regimes}}.
\newblock \emph{Annual Review of Statistics and Its Application}, 1\penalty0 (1):\penalty0 447--464, 2014.
\newblock \doi{10.1146/annurev-statistics-022513-115553}.

\bibitem[Wolpert(1992)]{wolpert1992stacked}
David~H Wolpert.
\newblock Stacked generalization.
\newblock \emph{Neural networks}, 5\penalty0 (2):\penalty0 241--259, 1992.

\bibitem[Mienye and Sun(2022)]{mienye2022survey}
Ibomoiye~Domor Mienye and Yanxia Sun.
\newblock A survey of ensemble learning: Concepts, algorithms, applications, and prospects.
\newblock \emph{IEEE Access}, 10:\penalty0 99129--99149, 2022.

\bibitem[Liu et~al.(2022)Liu, Zhong, Li, Seltzer, and Rudin]{liu2022fasterrisk}
Jiachang Liu, Chudi Zhong, Boxuan Li, Margo Seltzer, and Cynthia Rudin.
\newblock Fasterrisk: Fast and accurate interpretable risk scores.
\newblock \emph{Advances in Neural Information Processing Systems}, 35:\penalty0 17760--17773, 2022.

\bibitem[Hastie et~al.(2009)Hastie, Tibshirani, Friedman, Hastie, Tibshirani, and Friedman]{hastie2009additive}
Trevor Hastie, Robert Tibshirani, Jerome Friedman, Trevor Hastie, Robert Tibshirani, and Jerome Friedman.
\newblock Additive models, trees, and related methods.
\newblock \emph{The elements of statistical learning: data mining, inference, and prediction}, pages 295--336, 2009.

\bibitem[Shortreed et~al.(2014)Shortreed, Laber, Scott~Stroup, Pineau, and Murphy]{shortreed2014multiple}
Susan~M Shortreed, Eric Laber, T~Scott~Stroup, Joelle Pineau, and Susan~A Murphy.
\newblock A multiple imputation strategy for sequential multiple assignment randomized trials.
\newblock \emph{Statistics in medicine}, 33\penalty0 (24):\penalty0 4202--4214, 2014.

\bibitem[Nahum-Shani et~al.(2017)Nahum-Shani, Ertefaie, Lu, Lynch, McKay, Oslin, and Almirall]{nahum2017smart}
Inbal Nahum-Shani, Ashkan Ertefaie, Xi~Lu, Kevin~G Lynch, James~R McKay, David~W Oslin, and Daniel Almirall.
\newblock A smart data analysis method for constructing adaptive treatment strategies for substance use disorders.
\newblock \emph{Addiction}, 112\penalty0 (5):\penalty0 901--909, 2017.

\bibitem[Lu et~al.(2023)Lu, Howard, Gordon-Larsen, Meyer, Tien, Du, Wang, Zhang, and Kosorok]{lu2023constructing}
Minxin Lu, Annie~Green Howard, Penny Gordon-Larsen, Katie~A Meyer, Hsiao-Chuan Tien, Shufa Du, Huijun Wang, Bing Zhang, and Michael~R Kosorok.
\newblock Constructing a t-test for value function comparison of individualized treatment regimes in the presence of multiple imputation for missing data.
\newblock \emph{arXiv preprint arXiv:2312.15217}, 2023.

\bibitem[Shen et~al.(2023)Shen, Hubbard, and Linn]{shen2023estimation}
Jenny Shen, Rebecca~A Hubbard, and Kristin~A Linn.
\newblock Estimation and evaluation of individualized treatment rules following multiple imputation.
\newblock \emph{Statistics in Medicine}, 42\penalty0 (23):\penalty0 4236--4256, 2023.

\bibitem[Kahkoska et~al.(2023)Kahkoska, Freeman, Jones, Shirazi, Browder, Page, Sperger, Zikry, Yu, and {Busby-Whitehead}]{kahkoska2023Individualized}
Anna~R. Kahkoska, Nikki~LB Freeman, Emily~P. Jones, Daniela Shirazi, Sydney Browder, Annie Page, John Sperger, Tarek~M. Zikry, Fei Yu, and Jan {Busby-Whitehead}.
\newblock Individualized interventions and precision health: {{Lessons}} learned from a systematic review and implications for analytics-driven geriatric research.
\newblock \emph{Journal of the American Geriatrics Society}, 71\penalty0 (2):\penalty0 383--393, 2023.

\bibitem[Lorenzoni et~al.(2023)Lorenzoni, Petracci, Scarpi, Baldi, Gregori, and Nanni]{lorenzoni2023Use}
Giulia Lorenzoni, Elisabetta Petracci, Emanuela Scarpi, Ileana Baldi, Dario Gregori, and Oriana Nanni.
\newblock Use of {{Sequential Multiple Assignment Randomized Trials}} ({{SMARTs}}) in oncology: Systematic review of published studies.
\newblock \emph{British Journal of Cancer}, 128\penalty0 (7):\penalty0 1177--1188, March 2023.
\newblock ISSN 1532-1827.
\newblock \doi{10.1038/s41416-022-02110-z}.

\bibitem[Valiant(1984)]{valiant1984Theory}
Leslie~G Valiant.
\newblock A theory of the learnable.
\newblock \emph{Communications of the ACM}, 27\penalty0 (11):\penalty0 1134--1142, 1984.

\bibitem[McDonagh et~al.(2020)McDonagh, Selph, Buckley, Holmes, Mauer, Ramirez, Hsu, Dana, Fu, and Chou]{mcdonagh2020Nonopioid}
Marian~S. McDonagh, Shelley~S. Selph, David~I. Buckley, Rebecca~S. Holmes, Kimberly Mauer, Shaun Ramirez, Frances~C. Hsu, Tracy Dana, Rochelle Fu, and Roger Chou.
\newblock \emph{Nonopioid {{Pharmacologic Treatments}} for {{Chronic Pain}}}.
\newblock {{AHRQ Comparative Effectiveness Reviews}}. {Agency for Healthcare Research and Quality (US)}, {Rockville (MD)}, 2020.

\bibitem[Tibshirani(1996)]{tibshirani1996Regression}
Robert Tibshirani.
\newblock Regression {{Shrinkage}} and {{Selection Via}} the {{Lasso}}.
\newblock \emph{Journal of the Royal Statistical Society: Series B (Methodological)}, 58\penalty0 (1):\penalty0 267--288, 1996.
\newblock ISSN 2517-6161.
\newblock \doi{10.1111/j.2517-6161.1996.tb02080.x}.

\bibitem[{R Core Team}(2021)]{rcoreteam2021Language}
{R Core Team}.
\newblock \emph{R: {{A Language}} and {{Environment}} for {{Statistical Computing}}}.
\newblock {R Foundation for Statistical Computing}, {Vienna, Austria}, 2021.

\bibitem[Pocock and Simon(1975)]{pocock1975Sequential}
Stuart~J. Pocock and Richard Simon.
\newblock Sequential {{Treatment Assignment}} with {{Balancing}} for {{Prognostic Factors}} in the {{Controlled Clinical Trial}}.
\newblock \emph{Biometrics}, 31\penalty0 (1):\penalty0 103--115, 1975.
\newblock ISSN 0006-341X.
\newblock \doi{10.2307/2529712}.

\bibitem[Taves(1974)]{taves1974Minimization}
Donald~R. Taves.
\newblock Minimization: {{A}} new method of assigning patients to treatment and control groups.
\newblock \emph{Clinical Pharmacology \& Therapeutics}, 15\penalty0 (5):\penalty0 443--453, 1974.
\newblock ISSN 1532-6535.
\newblock \doi{10.1002/cpt1974155443}.

\end{thebibliography}

\clearpage

\appendix
\newgeometry{left=.75in, right=.75in}

\clearpage

\restoregeometry
\end{document}